\documentclass[12pt]{article}
\usepackage{doublespace,a4,epsfig,amsmath,amssymb,cite}

\begin{document}

\begin{center}
   \centering{\large\bf Specific heat upon aqueous unfolding of the protein interior: A theoretical approach}\\

  \vspace*{1cm}
  \centering{Audun Bakk\footnote{The author to whom correspondence should be addressed. \newline E-mail: Audun.Bakk@phys.ntnu.no}}, Johan S.\ H\o ye, and Alex Hansen \\
 {\it Department of Physics, Norwegian University of Science and 
     Technology, NTNU, NO-7491 Trondheim, Norway}\\
  
  \vspace*{0.5cm}
  \centering{(\today)}
\end{center}
\vspace*{0.5cm}

\begin{abstract}
We study theoretically the thermodynamics, over a broad temperature range (5$^{\circ}$C to 125 $^{\circ}$C), related to hydrated water upon protein unfolding. The hydration effect is modeled as interacting dipoles in an external field, mimicking the influence from the unfolded surfaces on the surrounding water compared to bulk water. The heat capacity change upon hydration is compared with experimental data from Privalov and Makhatadze on four different proteins: myoglobin, lysozyme, cytochrome c and ribonuclease. Despite the simplicity of the model, it yields good correspondence with experiments. With some interest we note that the effective coupling constants are the same for myoglobin, lysozyme, and cytochrome c, although they are slightly different for ribonuclease.  
\end{abstract}


\vspace*{0.5cm}

\section{Introduction}
\label{sec:1}
Proteins consist of 20 different amino acids with a great diversity with regard to size, polarity and charge. The understanding of water and interactions with water seems to be important in order to understand protein folding in general, and the special feature of cold unfolding of several small globular proteins in particular\cite{Privalov:86,Griko:88,Chen:89,Privalov:90,Franks:95,Graziano:97,Hansen:98,Los_Rios:00,Bakk:00}.

We in the present work will represent the energy difference between the unfolded and folded interior, with regard to the water, by mimicking additional hydrogen bonds from which we calculate the hydration heat capacity change upon protein unfolding. A justification of the model is the ability for water molecules to form an ``ice-like'' shell (``iceberg'' in the terminology of Frank and Evans\cite{Frank:45}) around apolar surfaces and thus create more hydrogen bonds. Reduction of both enthalpy\cite{Olofsson:84,Naghibi:86,Naghibi:87,Madan:97} and entropy\cite{Wilhelm:77,Dec:84} upon apolar hydration seems to be well established\cite{Makhatadze:95}. 

However, the protein interior that becomes hydrated upon unfolding, also consists of surfaces that has polarity, which means that the surface has permanent dipoles and charges. The heat capacity upon purely polar hydration becomes surprisingly negative\cite{Privalov:92a,Privalov:92b}. For apolar surfaces experiments show that the hydration contribution to the heat capacity upon solvation is positive. Also for proteins where part of the surface is polar this heat capacity is positive. Thus for simplicity we in this work will use the apolar ``ice-like'' shell picture to make a effective model for the hydration effect upon protein unfolding. In this way we may neglect some crucial features of polar solvation. 

 Finally, we apply equilibrium statistical mechanics to the model and calculate the hydration heat capacity increment, which we compare with experimental data from Privalov and Makhatadze on four different proteins\cite{Privalov:92b}.

\section{Hydration upon protein unfolding}
\label{sec:2}

We will use a refined version of a model first proposed by Hansen {\it et al.}\cite{Hansen:98,Hansen:99,Hansen:00}. The model studied here was applied by Bakk {\it et al.}\cite{Bakk:01e} on a complete protein folding model, but they did not study the hydration effect separately. In this work we will study specifically this hydration upon protein unfolding.

Protein unfolding involves a cavity formation in water with a rearrangement of the water molecules surrounding the unfolded protein\cite{Lee:85,Lee:91}. When estimating the solvation energy of exposing the interior of a protein to water, one has to calculate the energy difference between hydrated water, associated to the protein, and bulk water\cite{Privalov:92b}. More precisely, the hydration is defined as the transfer of a solute from a fixed position in the ideal gas phase to a fixed position in the solvent\cite{Ben-Naim:84}, i.e., water in the present case.

In order to model the effect upon inserting a surface into water, i.e., unfolding of a protein, we use the simplified analogy of a classical electrical dipole in an external electrical field  whose energy is   
\begin{equation}
   \label{eq:E_1}
   E_1=-\epsilon\cos{\vartheta},
\end{equation} 
where $\epsilon$ is a bending distortion constant. The angle $\vartheta$ is the polar angle. Eq.\ (\ref{eq:E_1}) is the hydration model used in the works by Bakk {\it et al.}\cite{Bakk:01e,Bakk:01c}, and it extends the interpretation of the hydration model applied by Hansen {\it et al.}\cite{Hansen:98,Hansen:99} and Bakk {\it et al.}\cite{Bakk:01a,Bakk:01b}.

The idea of representing the solvent by dipoles in protein folding was introduced by Warshel and Levitt\cite{Warshel:76}, and later applications by Russell and Warshel\cite{Russell:85}, Fan {\it et al.}\cite{Fan:99}, and Avbelj\cite{Avbelj:00}.

In addition to the energy due to the external field [Eq.\ (\ref{eq:E_1})] we will add a coupling term
\begin{equation}
   \label{eq:E_2}
   E_2=-\frac{1}{2}\sum_{i,j}J_{ij}\,{\bf s}_i\cdot{\bf s}_j,
\end{equation}
thus modeling pair interactions between the water molecules, where $J_{ij}$ is the coupling constant between water molecules $i$ and $j$, and ${\bf s}_i$ is the dipole moment of water molecule $i$. For simplicity we put $|{\bf s}_i|=1$. It can be shown that the energy $E_1+E_2$ [Eqs.\ (\ref{eq:E_1}) and (\ref{eq:E_2})] in a {\it mean field} solution\cite{Ma:80} can be represented as\cite{Hoye:80,Bakk:01e}
\begin{equation}
 E(\vartheta)=E_1+E_2=-(\epsilon+bm)\cos{\vartheta}
                                       +\frac{1}{2}bm^2,
\end{equation}
where $bm$, with $b=\sum_{j}J_{ij}$, is the mean field coupling between a water molecule and its surrounding water molecules, has to be determined self consistently. The average dipole moment is defined by $m=\langle cos{\vartheta}\rangle$. The standard mean field solution in an effective field $\epsilon_e=\epsilon+bm$ is
\begin{equation}
   \label{eq:m}
   m=\coth{\left(\frac{\epsilon_e}{RT}\right)}-\frac{RT}{\epsilon_e}.
\end{equation}
Note that here and below $\epsilon$ and $b$ are energies per mole, as the gas constant $R$ is used.

For $N$ dipoles per protein the partition function for the total hydration contribution upon protein unfolding is
\begin{equation}
   \begin{split}
   \label{eq:Z}
   Z&=\left[\int_0^{2\pi}d\varphi\int_0^{\pi}d\vartheta\,sin{\vartheta}
        \,\exp{\left(\frac{E(\vartheta)}{RT}
        \right)}\right]^{N}\\
      &=\left[\frac{4\pi RT\sinh{\boldsymbol{(}\epsilon_e/(RT)
               \boldsymbol{)}}}{\epsilon_e}\,\exp{\left(-\frac{bm^2}{2RT}
        \right)}\right]^{N}.
   \end{split}
\end{equation}
The total hydration heat capacity change per mole of proteins is 
\begin{equation}
   \label{eq:C}
   \Delta C=RT^2\frac{\partial^2}{\partial T^2}\ln{Z}.
\end{equation}
To obtain $\Delta C$, the self consistent Eq.\ (\ref{eq:m}) has to be solved numerically with respect to $m$. Eq.\ (\ref{eq:m}) can be obtained from Eq.\ (\ref{eq:Z}) via
\begin{equation}
   m=\frac{1}{N}\left(\frac{\partial\ln{Z}}{\partial\epsilon_e}\,
                      \frac{\partial\epsilon_e}{\partial\epsilon}
                      + \frac{\partial\ln{Z}}{\partial m}\,
                      \frac{\partial m}{\partial\epsilon}\right).
\end{equation} 
\section{Discussion}
\label{sec:3}
We want to compare the heat capacity change upon unfolding solvation of the protein interior with experiments. The proteins considered are myoglobin (Mb), lysozyme (Lys), cytochrome c (Cyt), and ribonuclease (Rns) which we compare with experimental data from Privalov and Makhatadze on the hydration contribution to the heat capacity change upon protein unfolding\cite{Privalov:92b}.

\begin{figure}
   \caption{}
   \vspace{1cm}
   \centering{\epsfig{figure=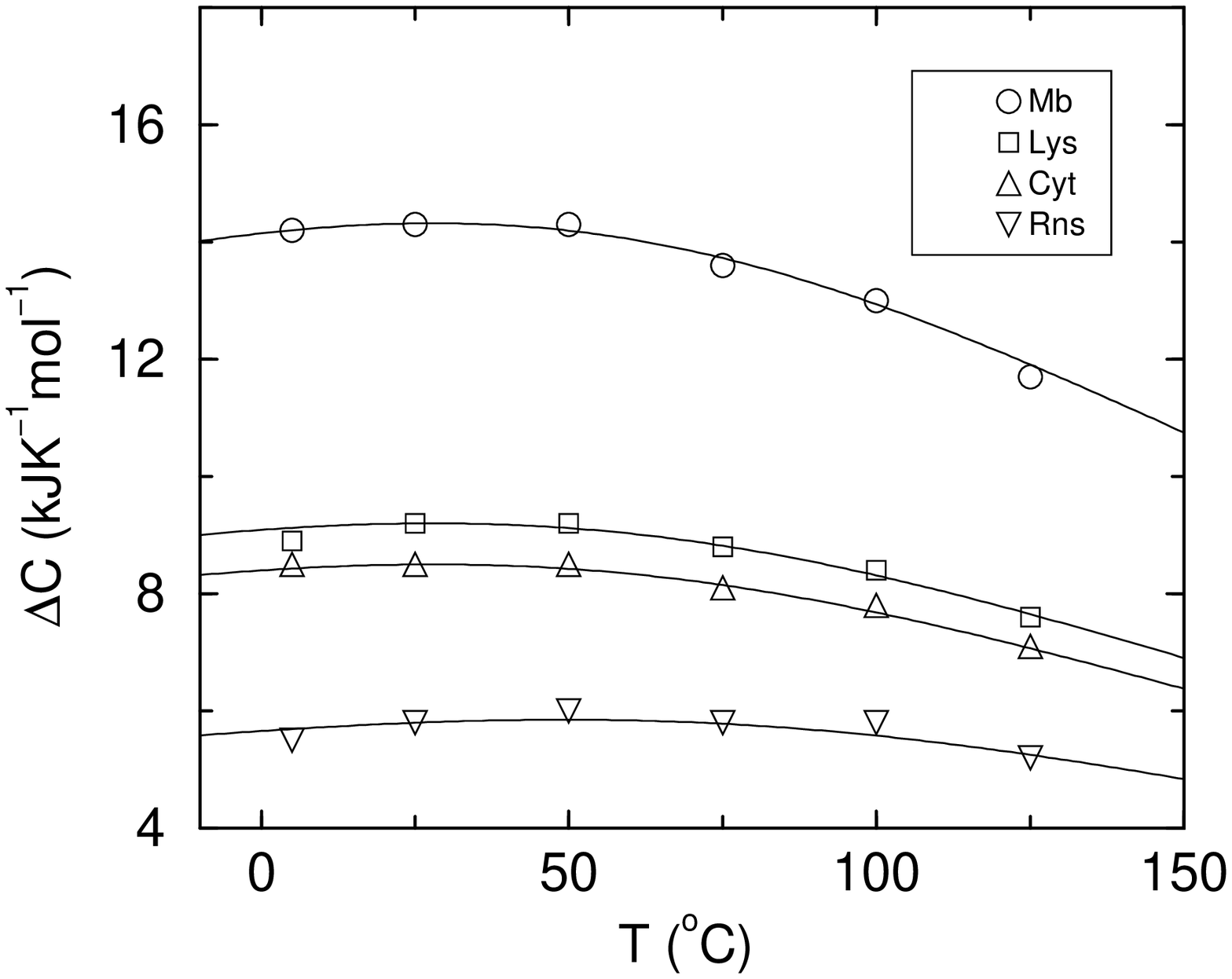,width=\linewidth}}
   \label{fig:1}
\end{figure}

The total hydration heat capacity change is shown in Figure \ref{fig:1}. The parameter fit to the experimental data agrees quite well with these data. 

Figure \ref{fig:1} shows that the heat capacity has a maximum around 25$^{\circ}$C for Mb, Lys, and Cyt, while this maximum is shifted to around 50$^{\circ}$C for Rns. Also in Table I a similar relation is reflected.
\begin{table}
\label{tab:1}
\begin{center}
\caption{Parameters, according to Eq.\ (\ref{eq:C}), used in Figure \ref{fig:1} for the fitting to the experimental hydration data from Privalov and Makhatadze\cite{Privalov:92b}. The difference in water accessible surface area between the unfolded and the folded protein $\Delta A_t$ is obtained from Makhatadze and Privalov\cite{Makhatadze:90}. $\Delta A_p/\Delta A_t$ is the ratio between the polar and total accessible surface area.} 
\vspace{0.3cm}

\centering{\begin{tabular}{c c c c c c}
              \hline {\bf protein} & $\epsilon$ & $b$  & $N$ 
                                   & $\Delta A_t$  & $\Delta A_p/\Delta A_t$\\
              & (kJmol$^{-1}$) & (kJmol$^{-1}$) & & (\AA$^2$) &(\%)\\
              \hline Mb  & 2.05 & 8.2 & 1240 & 18250  & 36.5\\
                     Lys & 2.05 & 8.2 & 800  & 14090  & 39.2\\
                     Cyt & 2.05 & 8.2 & 740  & 11830  & 38.2\\
                     Rns & 2.00 & 9.0 & 500  & 13300  & 44.8\\
              \hline
           \end{tabular}}
\end{center}
\end{table}
With some interest we note from Table I that both the ``electric field'' constant $\epsilon$ and the coupling constant $b$ are essentially the same for all of the four proteins, but there is a small deviation for Rns. 

This small deviation for Rns reflects itself in the ratios between accessible polar and total surface area (reported from Makhatadze and Privalov\cite{Makhatadze:90}), $\Delta A_p/\Delta A_t$, which are almost equivalent for Mb, Lys, and Cyt, while this ratio is significantly larger for Rns. Hence, the parameters $\epsilon$ and $b$ may be regarded as effective ones for the combined effect of apolar and polar surfaces as discussed in Section  \ref{sec:1}. 

One notes that Rns differ a bit from the other three proteins considered. This may reflect its larger fraction of polar surface which then also can affect qualitative properties. Our model is more like an effective one for a mixed polar and apolar surface. Thus features specific for polar surfaces are not properly taken into account, but are more or less taken into account by adjusting available parameters. E.g., our present model may seem to give too small curvature on figure 1 for Rns. A reason for this may be the negligence of quantization, which will lower the specific heat and thus increase its curvature for decreasing temperatures. The polar (ionic) forces are relatively strong and the hydrogen atom is light, which both favor quantum effects. Use of a two-level system\cite{Madan:96,Graziano:99,Bakk:01a} for this kind of problem can thus reflect such quantization. 

\section{Conclusion}
\label{sec:4}
We have proposed a thermodynamical model for the hydration of the protein interior that becomes exposed to water upon unfolding. To our knowledge this is the first model studied and compared to experimental data on the pure hydration heat capacity increment over such broad temperature range (5$^{\circ}$C to 125 $^{\circ}$C). 

Hydration is modeled in an ``ice-like'' shell analogy, where the water molecules are represented by interacting dipoles in an external field. Compared with experimental data from Privalov and Makhatadze\cite{Privalov:92b} for the four proteins myoglobin (Mb), lysozyme (Lys), cytochrome c (Cyt), and ribonuclease (Rns) the model fits quite well. The specific values of the the field coupling constant ($\epsilon$) and dipole coupling constant ($b$) [see Eq.\ (\ref{eq:Z})] are the same for Mb, Lys, and Cyt, while it is slightly different for Rns. 
 
In a future model one should take into account qualitative opposite effects, with respect to the heat capacity, by apolar and polar hydration. Experimentally one finds that the heat capacity change is negative for hydration of purely polar surfaces\cite{Privalov:92a,Privalov:92b}, in contrast to the apolar surfaces where this heat capacity change is positive. 

We have reason to expect that the present model, due to its simplicity, also may be useful in more complete protein folding models.

\section*{Acknowledgments}
A.\ B.\ thanks the Research Council of Norway for financial support (Contract No.\ 129619/410).  


\section*{Figure captions}
{\bf Fig. 1.}
The hydration heat capacity change upon unfolding of four different proteins. Experimental data from Privalov and Makhatadze\cite{Privalov:92b}. Parameters, corresponding to Eq.\ (\ref{eq:Z}), for the fit to the experimental data are listed in Table I.

\newpage

\vspace*{0.5cm}
\centering{A. Bakk,  A. Hansen, and J.\ S.\ H\o ye}\\
\centering{\it Specific heat upon aqueous unfolding of ...}

\end{document}